\documentclass[conf]{new-aiaa}
\usepackage{graphicx}
\usepackage[lofdepth,lotdepth]{subfig}
\usepackage[version=4]{mhchem}
\usepackage{siunitx}
\usepackage{xcolor}
\usepackage{epsf}
\usepackage{longtable,tabularx}
\usepackage[]{hyperref}
    \hypersetup{    colorlinks=true,
                    breaklinks=true,                
                    urlcolor= black,                
                    linkcolor= blue,                
                    bookmarksopen=false,
                    filecolor=black,
                    citecolor= magenta,
                    linkbordercolor=blue
}

\title{A Verification Framework for Certifying Learning-Based Safety-Critical Aviation Systems}

\author{Ali Baheri\footnote{Assistant Professor of Research, Department of Mechanical and Aerospace Engineering}}
\affil{West Virginia University, Morgantown, WV 26505}
\author{Hao Ren\footnote{Senior Advanced AI Engineer} and Benjamin Johnson\footnote{Senior Advanced AI Engineer}}
\affil{Honeywell Aerospace, Minneapolis, MN 55413}
\author{Pouria Razzaghi\footnote{Postdoctoral Researcher, Department of Mechanical and Aerospace Engineering} and Peng Wei\footnote{Assistant Professor, Department of Mechanical and Aerospace Engineering, AIAA Senior Member}}
\affil{George Washington University, Washington, DC 20052}

\begin{document}

\maketitle

\begin{abstract}
We present a safety verification framework for design-time and run-time assurance of learning-based components in aviation systems. Our proposed framework integrates two novel methodologies. From the design-time assurance perspective, we propose offline mixed-fidelity verification tools that incorporate knowledge from different levels of granularity in simulated environments. From the run-time assurance perspective, we propose reachability- and statistics-based online monitoring and safety guards for a learning-based decision-making model to complement the offline verification methods. This framework is designed to be loosely coupled among modules, allowing the individual modules to be developed using independent methodologies and techniques, under varying circumstances and with different tool access. The proposed framework offers feasible solutions for meeting system safety requirements at different stages throughout the system development and deployment cycle, enabling the continuous learning and assessment of the system product.
\end{abstract}

\section{Introduction}
%\subsection{Background}
\lettrine{A}lthough automation gives the ability to improve the performance and, in most instances, reduce the risk associated with a system, if it is not appropriately designed, thoroughly tested, and adequately verified it may pose a risk to people and property. As such, it is imperative that we understand the risks of automation and carefully consider these risks during design, development, testing, and implementation. With autonomous systems becoming more capable, they are entering safety-critical domains such as airborne systems and healthcare; ensuring the safe operation of these systems is a crucial step before they can be deployed. Failure to perform the proper degree of safety validation can risk the loss of property or even human life. Autonomy is increasingly implemented through artificial intelligence (AI) and machine learning (ML) applications, where, in most cases, it is tough to explain the inner workings of learning-enabled models and describe the behavior that can be understood with enough confidence for safe use in safety-critical systems.

AI has been demonstrated in recent research and development efforts such as Airbus’ Autonomous Taxi, Take-Off, and Landing (ATOL) project. The next-generation airborne collision avoidance system (ACAS X) is modeled as a Markov decision process, and its policy is compressed by a set of deep neural networks (ACAS Xu). Compared with the current avionics capability, these learning-enabled systems are expected to augment human pilots' perception and decision-making in complex and uncertain environments. Learning-enabled aviation systems are designed to handle uncertainty and probabilistic events as well as increase operational efficiency and scalability. However, aviation is a safety-critical application and it is critical to address the certification problem for these learning-based systems, especially the autonomous systems built with data-driven, blackbox-ish neural networks.

%Learning-enabled systems have unique characteristics that do not fit within existing FAA certification guidelines and policies, which are based on extensive testing and were not written with the needs and capabilities of autonomous learning-based systems in mind. The fact that these systems can modify their behavior at runtime contradicts certification guidelines in many ways. We argue that the significant technical barriers include, but are not limited to, (i) the need for comprehensive requirements. The dynamic nature of a learning-enabled system can make it difficult to specify the intended behavior precisely. (ii) a scarcity of feasible verification methods. Most learning-enabled systems are not amenable to verification by existing methods and would require an infeasible amount of testing to establish safety. (iii) nontransparent designs. Design information for a learning-enabled system significantly resides in the training data and not in the software implementation, making design evaluation impossible by traditional methods. We argue that future learning-based aviation systems will need to address these three overarching properties (detailed in \cite{NASA_OP}): Does a learning-based system possess the property of intent? Does a learning-based system possess the property of correctness? Does a learning-based system possess the property of innocuity?

Learning-enabled systems have unique characteristics that do not fit within existing FAA certification guidelines and policies, which are based on extensive testing and were not written with the needs and capabilities of autonomous learning-based systems in mind. The fact that these systems can modify their behavior at runtime contradicts certification guidelines in many ways. We argue that the significant technical barriers include, but are not limited to (i) the need for comprehensive requirements, (ii) a scarcity of feasible verification methods, and (iii) nontransparent designs. The dynamic nature of a learning-enabled systems can make it difficult to specify the intended behavior precisely. Even with a precise specification of the intended behavior, most learning-enabled systems are not amenable to verification by traditional methods, in part because their behavior is largely determined by training data rather than the software implementation; such methods would require an infeasible amount of testing to establish safety. We argue that future learning-based aviation systems will need to address three overarching properties: intent, correctness, and innocuity (detailed in \cite{NASA_OP}).

\begin{figure}[t]
  \centering
    \includegraphics[width=0.6\textwidth]{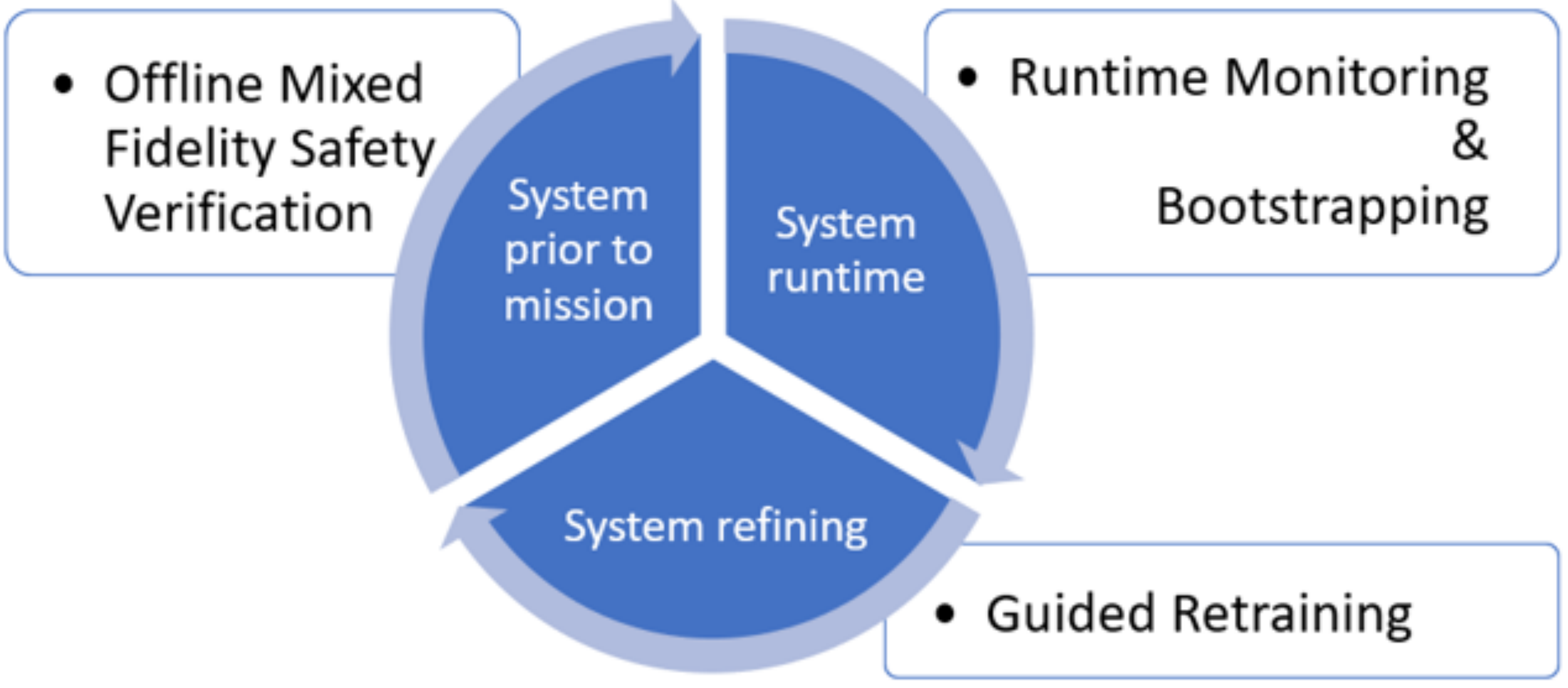}
    \caption{Safety verification frameworks throughout learning-enabled system mission cycle.}
    \label{fig:cycle}
\end{figure}

In this paper, we propose a safety verification framework to address these questions throughout the mission cycle of the learning-based system, as shown in Fig.~\ref{fig:cycle}. Specifically, the offline multi-fidelity safety verification will check the property of intent and innocuity specified in formal languages to the best extent before system execution. Secondly, the proposed online monitoring module will to monitor the property of innocuity in runtime to avoid undesired behavior, which the verification may not capture prior to execution due to environmental uncertainty.
%Last but not least, the bootstrap safeguard will collect critical statistics from possible innocuity and/or its precursors closely related to out-of-sample scenarios that are then used to refine the system design or augment the learning-based component training.
Last but not least, the bootstrap safeguard augment the innocuity property by collecting critical statistics about possible behaviors related to out-of-sample scenarios that are used to refine the system design or augment the learning-based component training.

%%%
%%%

\begin{figure}[t]
\centerline{\includegraphics[width=0.8\textwidth]{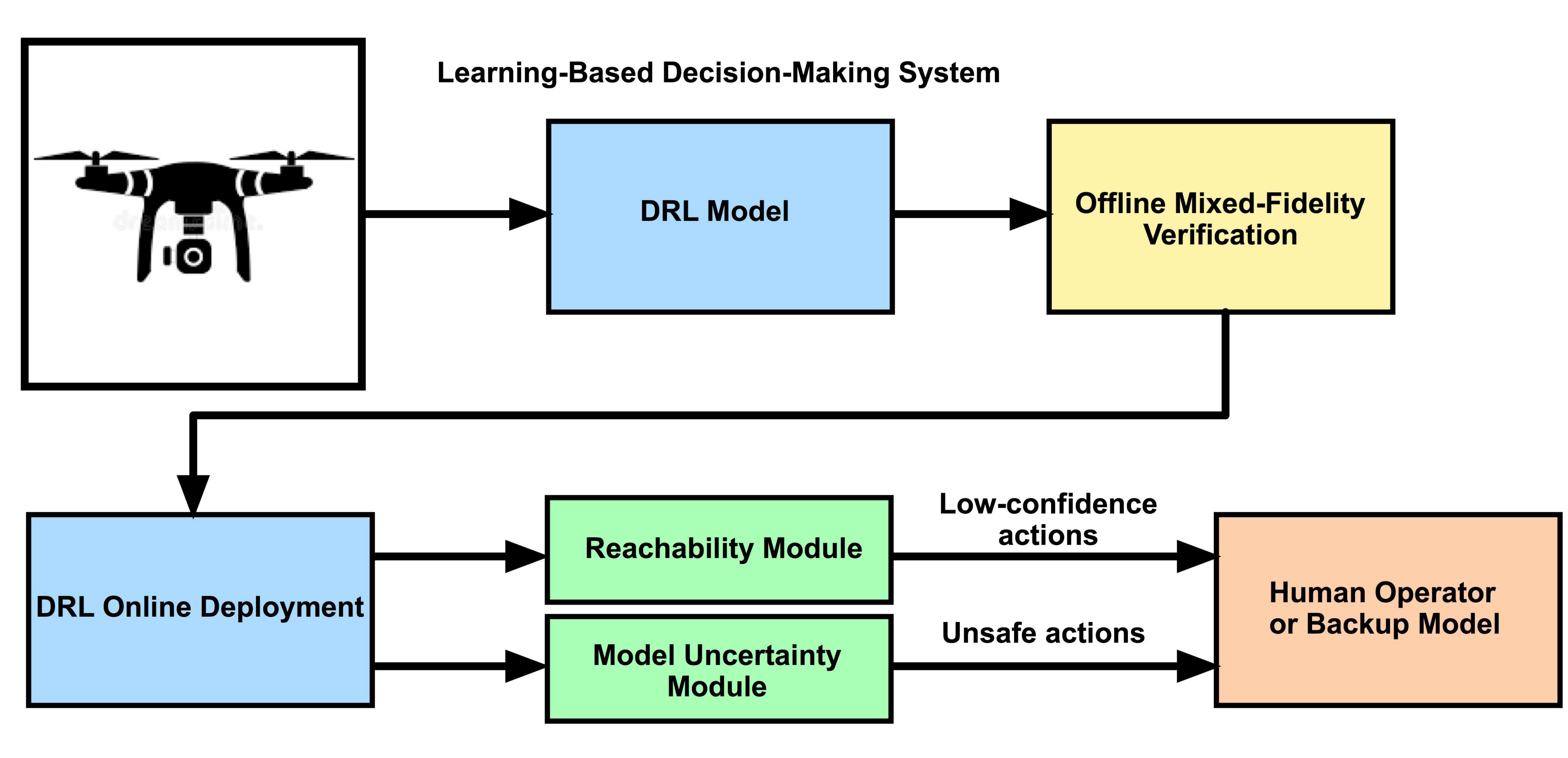}}
\caption{The integration of the proposed safety validation framework for learning-based safety-critical aviation systems. The blue blocks are learning-based aviation system model development and implementation. The yellow block is the offline verification tool. The green blocks are two online verification modules integrated as the online safeguard.}
\label{fig:framework}
\end{figure}

%We seek to answer the following research question: can we design and implement a safety verification framework for a complex or learning-based aviation system? 
Our top-level goal is to formally and continuously verify a given learning-based system's design and mission safety with offline mixed-fidelity optimization and online safeguard monitoring and bootstrapping. To achieve this goal, we propose a holistic, integrated verification framework with two thrusts shown in Fig.~\ref{fig:framework}. In this position paper, we briefly introduce a safety framework aimed at addressing the barriers mentioned in (ii) and (iii) through offline safety verification, runtime monitoring, and bootstrapping, as detailed in Sec.~\ref{sec:offline}, Sec.~\ref{sec:onlinertm}, and Sec.~\ref{sec:onlineboot}.  

%\subsection{Goals}
%The benefits and impacts of this project include: (1) the proposed safety framework offers feasible solutions for safety verification of learning-based decision-making models in aviation systems; and (2) it provides a viable path to certify the learning-based components in avionics systems.

\section{Background and Related Work}

\paragraph{Offline verification.} 

%Many studies used the formal methods for safety verification of safety-critical autonomous systems. To examine the system's reliability under test and fulfill the safety specifications, different mathematical and computational models have been conducted. The critical challenge to employing formal methods for safety verification is the scalability issues. Learning-based components in avionics systems typically have a large, continuous, and high-dimensional design space, which becomes intractable when applying formal methods-based safety verification. Researchers in the verification and validation community have recently been interested in black-box safety validation methods. Black-box stress testing comes with the benefit that sometimes organizations are not interested in revealing some internal aspects of the system under test. However, they still want to have a notion of safety for the system.

Formal methods have been widely used for the safety verification of safety-critical autonomous systems. These methods build a rigorous mathematical or computational model of the system under test to reliably verify the system or exhaustively examine whether a set of safety specifications have been fulfilled or not. Formal methods include various techniques, ranging from model checking to automated theorem proving, among others. Model-checking approaches formally check whether a finite-state model of a system fulfills a given specification \cite{clarke2018handbook}. In a similar vein, automated theorem proving leverages computer algorithms to generate mathematical proofs automatically \cite{gallier2015logic}. The key challenge to employing formal methods for safety verification of complex systems is the curse of dimensionality. Autonomous decision-making models in avionics systems in general consist of a large, continuous, and high-dimensional design space, making them intractable for formal verification methods. In another line of work, black-box and gray-box safety validation tools have recently gained interest in the verification and validation community. These techniques include planning algorithms \cite{tuncali2019rapidly,dreossi2015efficient}, a number of optimization tools \cite{silvetti2017active,deshmukh2017testing,deshmukh2015stochastic,mathesen2019falsification}, and reinforcement learning \cite{lee2015adaptive,akazaki2018falsification}.

%The literature is replete with black-box safety verification approaches, including genetic algorithm \cite{srivastava2009application, zou2014safety}, genetic programming \cite{corso2020interpretable}, simulated annealing \cite{kapinski2015simulation}, extended ant colony optimization \cite{annapureddy2010ant, annpureddy2011s}, stochastic local search \cite{deshmukh2015stochastic}, cross entropy method \cite{sankaranarayanan2012falsification,kim2016improving}, and Bayesian optimization \cite{silvetti2017active,deshmukh2017testing,ghosh2018verifying,abeysirigoonawardena2019generating}, among many others. In black-box safety validation, the task is framed as a black-box global optimization problem where the decision variables are the test inputs (system parameters and environmental parameters) and the objective function (metrics to define failure modes) is to find failure scenarios.

\paragraph{Runtime monitoring.} Runtime monitoring  requires the specification of properties that the software or system must satisfy and monitoring those properties during execution to ensure that they are met. Should the system violate the specified properties during execution, the runtime monitoring module must rapidly and accurately detect the violation and indicate it as such (the indication may take the form of an alert to an operator or may result in an automatic adjustment of the system behavior). One popular choice to specify the safety properties is to use temporal logic, such as Signal Temporal Logic (STL) \cite{maler2004monitoring}, due to its rich semantics in describing a broad range of real-valued, temporal properties in cyber-physical systems. Furthermore, an STL formula can be evaluated to provide a measure of the “robustness” with which the system satisfies (resp. violates) the STL specifications by evaluating how close the system is to violating (resp. satisfying) the formula \cite{deshmukh2017robust}. When well designed, a runtime monitoring module can be used to detect, predict, and/or avoid unsafe or undesired behavior during the execution of a complex cyber-physical system. In order to predict or avoid unsafe behavior, reachability analysis given environmental uncertainty can be used to compute an estimate of the set of states the system might reach in bounded time. For safety verification purposes, a common approach involves computing the reachable set over-approximation symbolically \cite{frehse2011spaceex,lin2020reachflow} from a finite set of the initial state. In previous research \cite{ren2019simulation}, we studied a simulation-based verification approach that simulates a finite set of forwarding traces from within the initial state set and performs over-approximations around those, thereby covering the initial states that were not simulated and relying on specific dynamic properties such as Lipschitz continuity.

\paragraph{Safe reinforcement learning.} There has been a significant amount of research on safe (deep) reinforcement learning for a variety of applications, ranging from driverless cars to robotics to aviation systems \cite{garcia2015comprehensive,isele2018safe,baheri2020deep,baheri2022safe,guo2021safety}. In addition, there are other online methods to enhance the safety of deep reinforcement learning (DRL) models by dealing with model and state uncertainties and rejecting existing noise. For safeguarding DRL models, an online certified defense is designed in \cite{lutjens2020certified}. It computes guaranteed lower bounds on state-action values to identify and choose the optimal action under a worst-case deviation in input space due to possible adversaries or noise. A safe reinforcement learning framework \cite{lutjens2019safe} uses bootstrapping to provide computationally tractable uncertainty estimates to bridge the gap between training scenarios and testing scenarios. The MC-dropout methods have been used to estimate the model uncertainties \cite{lutjens2019safe, gal2016improving, kahn2017uncertainty, malik2019calibrated}. Additionally, the data augmentation method is used to reject the state uncertainties generated by noise from sensors, communication, or observations \cite{cobbe2019quantifying, krizhevsky2012imagenet, laskin2020reinforcement}. The preemptive shields and post-posed shields are synthesized to enforce runtime safety for reinforcement learning agents \cite{alshiekh2018safe}. 
%The efficacy of the framework has been demonstrated in multi-agent systems.

\footnotetext{We use gym-PyBullet-drones \cite{panerati2021learning} as our main simulator platform. PyBullet provides a suite of Gym environments, a must-have for our UAV controller development and verification, and a convenient interface for rapidly deploying and testing our safety verification frameworks. PyBullet has not yet shown support for realistic environment rendering, but it could customize and render basic shapes/objects relatively easily for our algorithms' correctness demonstration.}

\section{Methods}
\subsection{Offline Safety Verification}
\label{sec:offline}

The ultimate objective of this work is to propose a rigorous notion of assurance in learning-enabled safety-critical aviation systems. This assurance could be built using some offline or online verification processes. The drawback of online verification methods is that they require some form of real-world deployment, which could be unsafe and risky. Typically, it is of great interest to reveal possible failure scenarios in a simulated environment before deploying a learning-enabled system into the real world. Since the space of failure events and edge cases is vast in complex systems, the validation process might be very time-consuming as many computationally expensive simulations are required for safety validation. We argue that a system under test could query data from multiple sources, including different levels of fidelity in simulated environments. 

The level of fidelity specifies the degree to which a simulator takes into account simplifications and assumptions when modeling a system. Low-fidelity simulators make strong assumptions and simplify the underlying system, enabling relatively fast execution. On the other hand, high-fidelity simulators make mild assumptions about the underlying system and show more realistic behaviors and dynamics. However, they are much slower to execute than low-fidelity simulators. Currently, there is no rigorous mechanism to reason about the safety behaviors of a learning-enabled decision-making system that (optimally) takes into account data from mixed-fidelity learning environments. At this point, systematic approaches that capture the information from multiple simulated environments could significantly speed up the certification process and reduce the overall computational time and cost. In this position paper, we outline two approaches for multi-fidelity verification of learning-based decision-making systems: (i) multi-fidelity Bayesian optimization and (ii) multi-fidelity reinforcement learning.

%\begin{itemize}
%    \item \ul{Multi-fidelity Bayesian optimization for falsification task:} Given a learning-enabled decision-making system, we aim to discover instances from the uncertainty space that violate the safety specification rules from multiple sources of information \cite{baheri2017real}.
%    \item \ul{Multi-fidelity reinforcement learning for most likely failure analysis:} It is also of interest to identify \emph{most likely failure scenarios} hinged upon some measure of likelihood of occurrence before deployment of an AI-enabled system into the real world. We seek to utilize multi-fidelity reinforcement learning algorithm aimed at finding a policy that maps states to disturbances causing the system to fail.
%\end{itemize}

%The key limitation of the existing approaches is they are slow and very sample-inefficient. To speed up the certification process, we propose an offline safety verification framework to coordinate and integrate knowledge from various-fidelity testing environments. We develop rigorous mathematical guarantees and proofs to set up experiments across multiple fidelity levels by combining formal methods and multi-fidelity optimization techniques \cite{beard2022safety}.

\paragraph{Multi-fidelity Bayesian optimization approach.} Bayesian optimization is a statistical machine learning technique that seeks to find the global optimum of a black-box function within only a few function evaluations, as each function evaluation is typically expensive. Bayesian optimization has been applied to various real-world problems, ranging from real-time control \cite{baheri2017real} to falsifying a cyber-physical system \cite{deshmukh2017testing}. Given a system under test and a safety specification, we propose a multi-fidelity Bayesian optimization to falsify a system that could query multiple sources of information (e.g., a multi-fidelity simulated environment). Specifically, our objective is to arrive at a principled algorithmic falsification framework that uses multi-fidelity simulated environments. Multi-fidelity Bayesian optimization provides a mechanism to decide \emph{when} and \emph{where} to evaluate the high-fidelity model. More precisely, we cast the falsification problem as a sequential decision-making problem using Bayesian optimization, where at each iteration, we not only search over the space of possible disturbance trajectories that cause the system to fail but also whether to evaluate a low-fidelity or high-fidelity simulated environment. 

\paragraph{Multi-fidelity reinforcement learning approach.} It is also of interest to leverage the sequential structure of the problem at hand and find the most likely failure events. In a multi-fidelity reinforcement learning setting, the black-box verification problem is treated as finding the most likely failure paths based upon some metric of the likelihood of occurrence. The proposed multi-fidelity reinforcement learning approach is built on the concept of adaptive stress testing (AST) for testing a complex decision-making system where the goal is to find a policy that maps states to disturbances causing the system to fail \cite{lee2015adaptive}. The failure scenario is defined as the sequence of actions that results in a failure event. One key limitation of AST is that it fails to find particularly sparse failures and can be inclined to find similar solutions to those found previously. To overcome this challenge, we leverage multi-fidelity learning to lessen the overuse of information. That is, information in lower fidelity simulations can be used to build up samples less expensively, and more effectively cover the solution space to find a broader set of failures. 

Our ultimate goal is to reduce the number of samples from high-fidelity simulators. Inspired by recent reinforcement learning approaches that use multiple fidelities of simulators and integrate information from each \cite{cutler2015real}, we propose a mechanism that uses information from a low-fidelity environment to guide search in a high-fidelity environment and updates the learned model of the low-fidelity model using information from the high-fidelity environment. Our preliminary results evaluated in a grid world environment demonstrate that this approach results in a lower number of samples from high-fidelity environments while maximizing the number of discovered failures \cite{beard2022safety}.

%%%
\subsection{Runtime Safety Monitoring}
\label{sec:onlinertm}
Since the learning-based components are data-driven, they can usually generalize and perform well in the test scenarios, which have similar distributions to the training scenarios. However, after these learning-based components are deployed in aviation systems, they will face new scenarios that have never been seen during their training/learning stage. The model will not feel "confident" in these out-of-sample scenarios and will provide untrustworthy actions. We expect the offline verification tools developed in Sec.~\ref{sec:offline} will capture most of the unsafe actions recommended by the learning-based components. However, online safety monitoring is still necessary to detect dangerous actions or decisions during flight operations.

We will develop a runtime safety monitoring module that predicts likely vehicle trajectories in bounded time based on the vehicle's dynamics and environmental conditions and evaluates its safety against a set of properties, specified as STL formulas. The safety monitoring module is designed to detect potential unsafe actions recommended by the learning-based components in runtime. When the recommended action/decision exhibits precursor behaviors that may potentially cause the vehicle trajectory to reach an unsafe state, violating the one or more safety specifications, the monitoring module will alert the vehicle control (or switch to a backup controller) to resort to a safe action. Assuming that the runtime monitoring module correctly identifies violations of the safety constraints before their violation, we can switch to a backup controller that verifiably avoids any such violation (possibly with performance trade-offs), or a safe-fail mechanism. In that case, we can guarantee the overall safety of the vehicle. In this paper, we consider safety against only static geo-fence and static environmental obstacles within (e.g., electricity poles), as well as pre-defined controllable intruder vehicles. Safety violation caused by a neither controllable nor predictable intruder vehicle is an exciting topic for further research, but is out of the scope of this work. However, our work can be extended to multi-agent operations with all agents having complete control access.

\begin{figure}[t!]
\centerline{\includegraphics[width=0.9\textwidth]{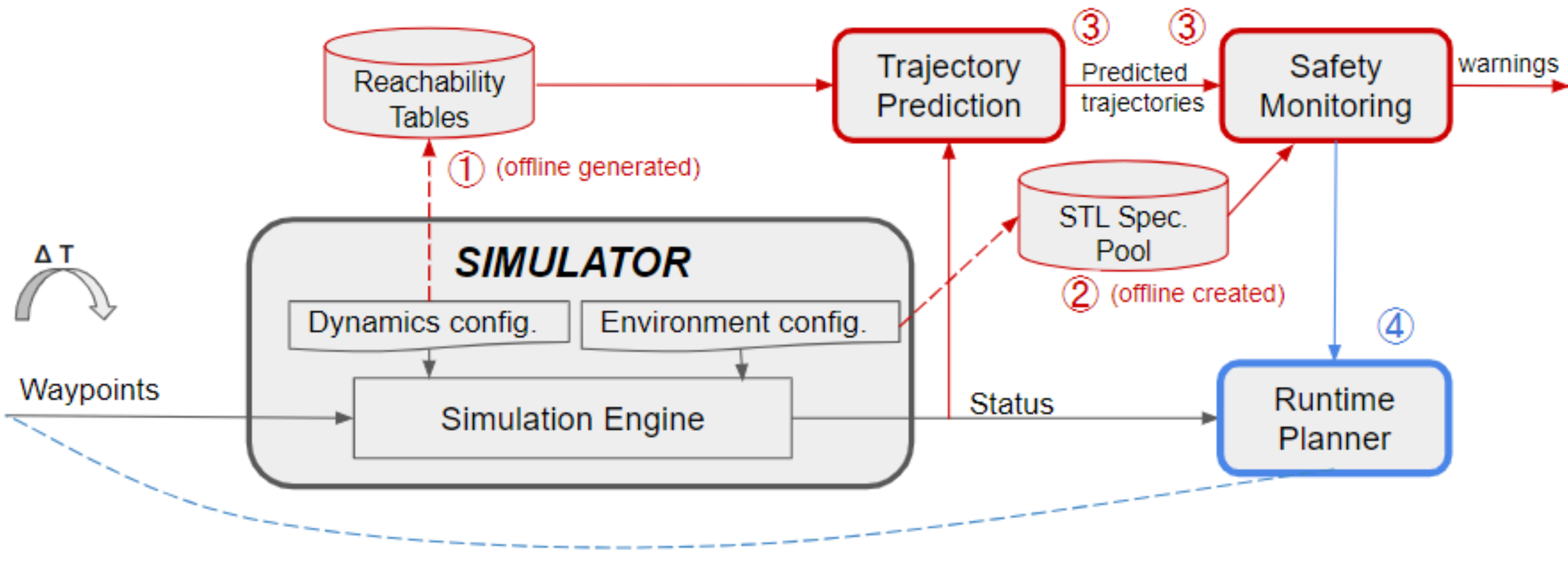}}
\caption{Reachability-based trajectory prediction and safety monitoring overview and development tasks: \textcircled{1} generating the reachability tables offline; \textcircled{2} creating STL safe guards specification pool offline; \textcircled{3} integrating safety monitoring with trajectory prediction; \textcircled{4} closing the verification loop for runtime planning.}
\label{fig:reach}
\end{figure}

Fig.~\ref{fig:reach} breaks down the module's components built around the simulator and development tasks. In the typical execution of a pre-planned UAV mission simulation (depicted by the box and line elements in black), the simulator engine uses waypoints provided by the planner to generate a set of control signals for the individual rotors based on the dynamics configuration information, and outputs the vehicle status after applying the controls and the interaction with environmental factors (wind, obstacle, etc.). Such execution updates every $\Delta T$ time. The safety monitoring module (depicted by the box and line elements in red) interacts with the simulator at the same update frequency. Firstly, due to the extensive computation cost for reachability analysis, a set of reachability lookup tables are pre-computed offline (marked by a dotted line), containing the bounded time, over-approximated position displacement corresponding to different intervals of velocity and acceleration. This allows the runtime trajectory prediction component to quickly calculate an over-approximation of the possible vehicle trajectories via the pre-computed tables, given its current state. Secondly, an STL specification pool is created manually offline (marked by a dotted line) with the desired safety requirements and environmental constraints. The third component evaluates the satisfaction of the STL specifications to monitor the execution of the system. The resulting safety margins will be displayed via a graphical interface, which will publish a set of warnings for potential violations and assist in bootstrapping data collection. We also envision that, in the future, the safety monitoring module can be integrated with the runtime planner (depicted in a blue box and line elements) so that the reachable states of the vehicle can be fed back to the deep neural net controller, to predict the reaction maneuver further, forming a closed-loop verification for the entire mission period.

\begin{figure}[b]
\centerline{\includegraphics[width=0.9\textwidth]{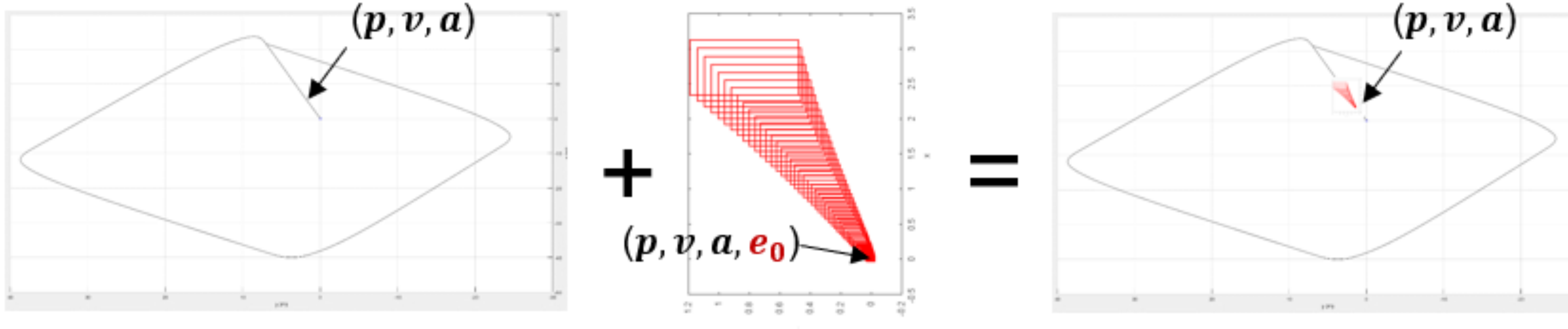}}
\caption{Adding the position change overapproximation (reachable tube shown in red rectangles) onto runtime trajectory (shown in gray trace) in $2$-D simulation. Each of the red rectangles indicate the reachable set of position between two consecutive time frame used in analysis. $p$: position; $v$: velocity; $a$: acceleration; $e_0$: initial uncertainty. }
\label{fig:comp}
\end{figure}

\paragraph{Reachability-based trajectory prediction.} Simulation is often used for trajectory prediction. Due to simulation error and environmental uncertainties, single or finite simulated trajectories are not sufficient to prove vehicle safety. In our previous study, we developed a formal approach for over-approximating system reachable states from finite simulations for safety purposes \cite{ren2019simulation}. Our approach uses an on-the-fly dynamic repartitioning scheme for the bounded-time reachability of hybrid dynamics, achieving less over-approximation error accumulation (even constant bounded for convergent dynamics), a critical premise for practical uses. Under this approach, over-approximation errors grow at different rates according to system dynamics and parameter configurations. This limits the prediction period under specific precision requirements. For trajectory monitoring, given bounded ranges of initial vehicle state (position, velocity, acceleration, etc.) and bounded environmental uncertainty factors (wind, for example), the reachability algorithms can compute offline the reachable set that contains all possible trajectories up to (typically) $5$-$20$ seconds in the future. Such reachability analysis is too computationally costly and slow to be applied at runtime. Instead, the bounded time reachable set can be decomposed based on relative position, since the position change depends only on vehicle dynamics. We then generate lookup tables that cover all the possible vehicle states related to a relative position by partitioning their ranges into subrange cells. The position change that corresponds to each subrange cell is computed offline while satisfying the precision requirement. Then, for runtime trajectory prediction, the runtime position from the onboard navigation module is used as the starting position. The position change for a given time span is obtained from the lookup tables by checking which cell the runtime states fall into. Their addition gives the comprehensive monitoring information shown in Fig.~\ref{fig:comp}. For dynamics whose dimensions are independent of each other (such as in our case), the lookup tables can be computed in $1$-D form and quickly applied to $2$-D and $3$-D scenarios. 
  %Since the position change is computed from ranges of states instead of specific values of states, it introduces a new source of over-approximation error that leads to the trade-off between precision and lookup table size. 
When only the current time's control (e.g., thrust, rotor speeds, acceleration) is available, the monitoring module can either assume the control will remain for the duration of the prediction period or assume no control at all future times. The former corresponds to constant control in circular hovering or uniform acceleration scenarios. In contrast, the latter corresponds to the end of a control period or a sudden loss of control. Further, predictions can be made to accommodate the future controls when the nominal plan is available to the monitor. All three of the above types will be implemented and used in different STL safety specifications.

\begin{figure}[t]
\centerline{\includegraphics[width=0.8\textwidth]{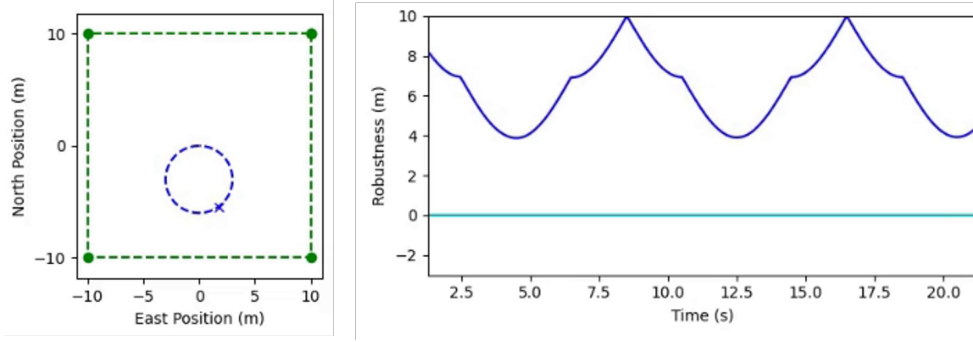}}
\caption{An example of the robustness value (in blue, right) of an STL formula over time, for a UAV that is required to remain within a geofenced region (green, left) while following a predetermined path (blue, left).}
\label{fig:stl}
\end{figure}

% The STL provides a formal framework for defining the properties as a set of logical conditions and constraints on the state of the system (i.e., the vehicle and its environment). In particular, STL allows for the rigorous specification of complex temporal properties of one or more continuous states of the system; for example, one might define a formula that requires that the vehicle stay within a set boundary area unless doing so would probably cause a collision with a static or dynamic obstacle. Once properly defined, such properties can then be monitored during system execution by evaluating their satisfaction over a partial trace of the system (note that a full trace of the system would require the complete execution, which is unavailable).
\paragraph{STL-based safe guards.} In runtime monitoring, safety requirements, constraints, and nominal behaviors are expressed as logical formulas over the system's past, current, and/or possible future states. We use STL to define our safety properties \cite{maler2004monitoring}, which provides a formal syntax and semantics for specifying properties of a system as logical formulas over the time-varying state of the system, including formally-defined methods for evaluating their satisfaction. In particular, STL allows for the rigorous specification of complex temporal properties of one or more continuous states of the system; for example, one might define a formula that requires that the vehicle stay within a set boundary area unless doing so would probably cause a collision with a static or dynamic obstacle. Once properly defined, such properties can then be monitored during system execution by evaluating their satisfaction over a partial trace of the system (note that a full trace of the system would require the complete execution, which is unavailable). Importantly, techniques exist to perform online monitoring of a system with STL specifications \cite{deshmukh2017robust}. Such techniques provide both an indication of whether the specifications are satisfied or not (i.e., their Boolean truth values) as well as a measure of how close the system is to violating the specification (i.e., the robustness with which the continuous state satisfies the specification), and do so formally and rigorously. Figure~\ref{fig:stl} shows, on the right, a plot of the robustness measure over time for an STL formula that requires that a UAV remain within a specified region while following a predefined path; the robustness value (solid blue line) indicates how close the UAV approaches the geofenced boundary, with positive values indicating that it remains inside the boundary (satisfying the specification, in a Boolean manner). The left-side figure in Fig.~\ref{fig:stl} depicts a $2$-D view of the position of the UAV (the blue X), the path of the UAV (dashed blue line), and the assigned geofence (dashed green line).
\subsection{Runtime Bootstrapping}
\label{sec:onlineboot}
The learning-based aviation system under test is a DRL model presented in \cite{brittain2021autonomous, guo2021safety}, which was designed for aircraft separation assurance. In this proposed DRL framework, individual aircraft are represented by agents in en-route airspace, allowing autonomous separation from other aircraft. Using sequential decisions in real-time, agents will assess surrounding air traffic conditions and select speed advisories in real-time to avoid conflict at route intersections and along each flight route. Due to the decentralized setting, the framework's complexity remains the same for a variable number of aircraft and intersecting paths in an airspace sector. 
%Meanwhile, throughout the mission, the statistics-based module collects and analyzes the controller data, detecting significant variance (out-of-sample decision confidence level) and/or any other statistical abnormalities for post-mission root-cause analysis and refinement guidance.

\begin{figure}[ht!]
\centerline{\includegraphics[width=0.6\textwidth]{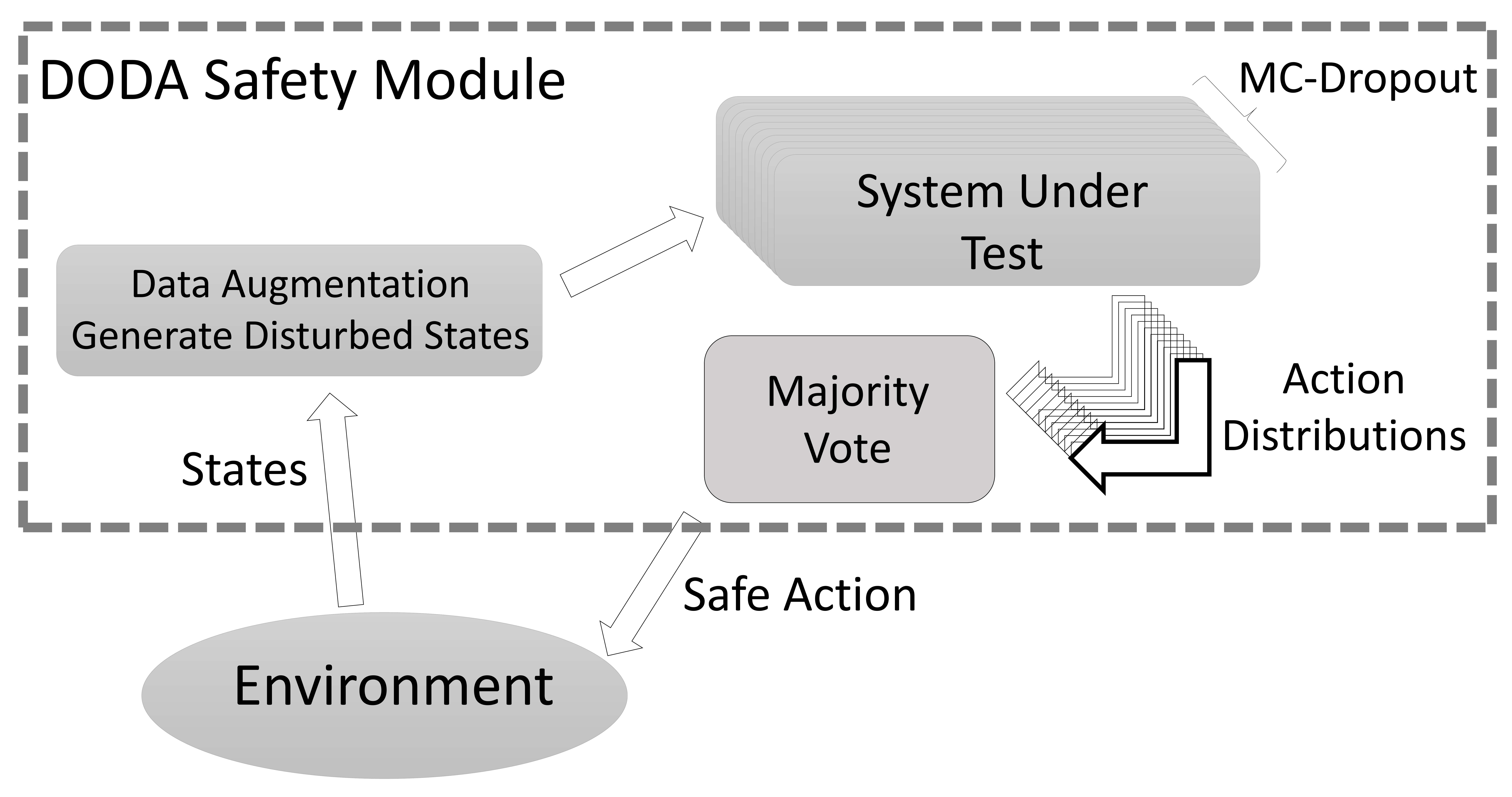}}
\caption{Structure of DODA safety module \cite{guo2021safety}. An agent observes the environment and selects the safe action. At each
time step, disturbed states are generated. Each disturbed state is used as the input state to the DRL model for action selection. An MC-dropout ensemble fits the action distribution for each disturbed state. Then, disturbed states and the corresponding action distributions are used in the DA module. Majority vote criterion is used to select the final action.}
\label{fig:DODA}
\end{figure}

By implementing two sub-modules for the DRL model, we will provide immediate safety enhancement for the DRL agent in autonomous separation assurance. We will develop an online bootstrapping method to estimate the variance of the recommended action under the current scenario. More concretely, we propose to implement two bootstrapping mechanisms: MC-dropout and data augmentation (DODA). These two mechanisms will provide computationally tractable uncertainty estimates, by which we can avoid low-confidence decisions and select the safest decision. Specifically, we will implement a model safety sub-module to assess the uncertainty of the DRL model based on MC-dropout and a state safety sub-module that incorporates execution-time data augmentation to evaluate the uncertainty of the state observations as illustrated in Fig.~\ref{fig:DODA}.

\begin{figure}[h!]
    \centering
   \subfloat[]{{\includegraphics[width=0.45\textwidth]{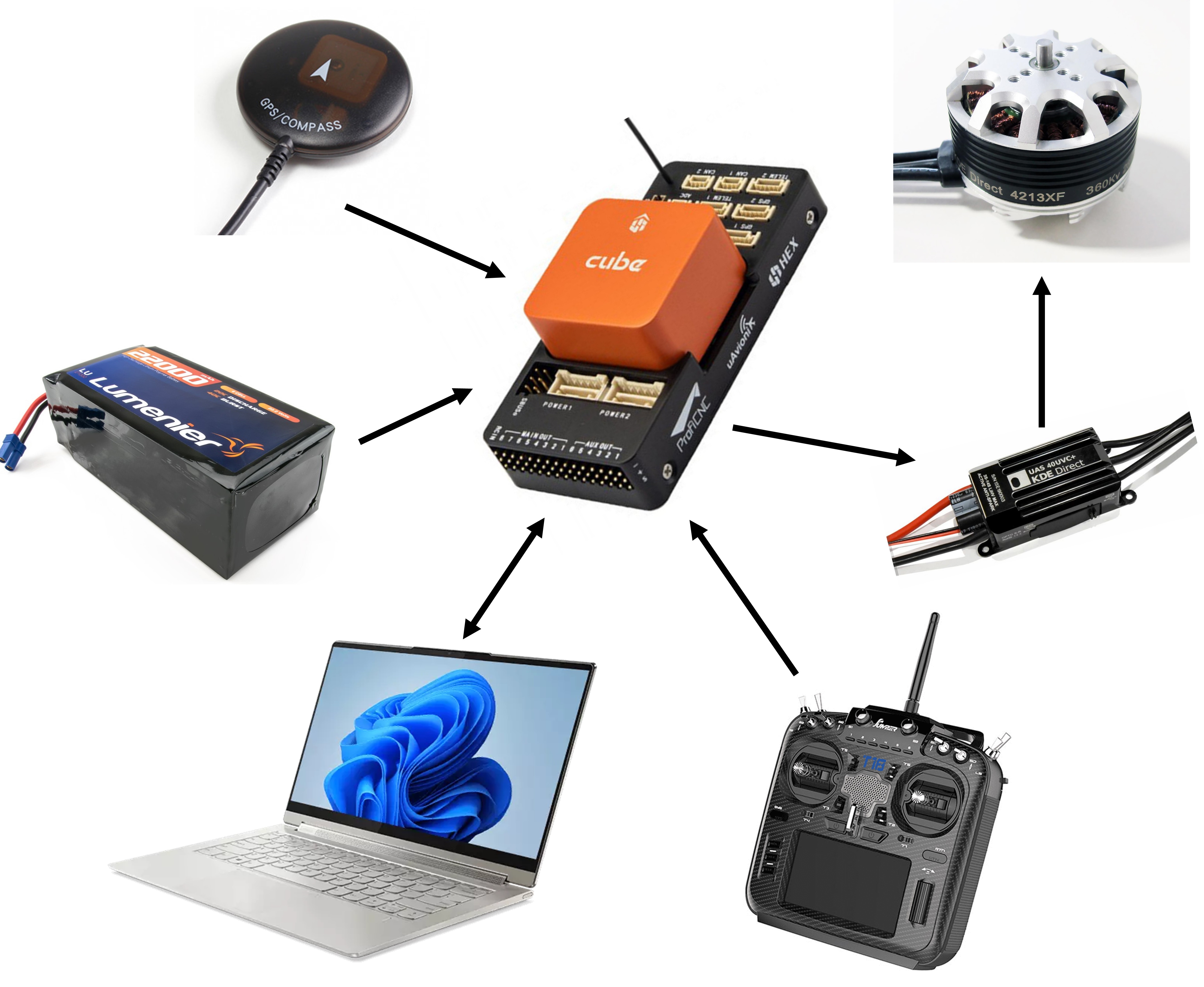}\label{fig:SetupScheme} }}
    \qquad
   \subfloat[]{{\includegraphics[width=0.45\textwidth]{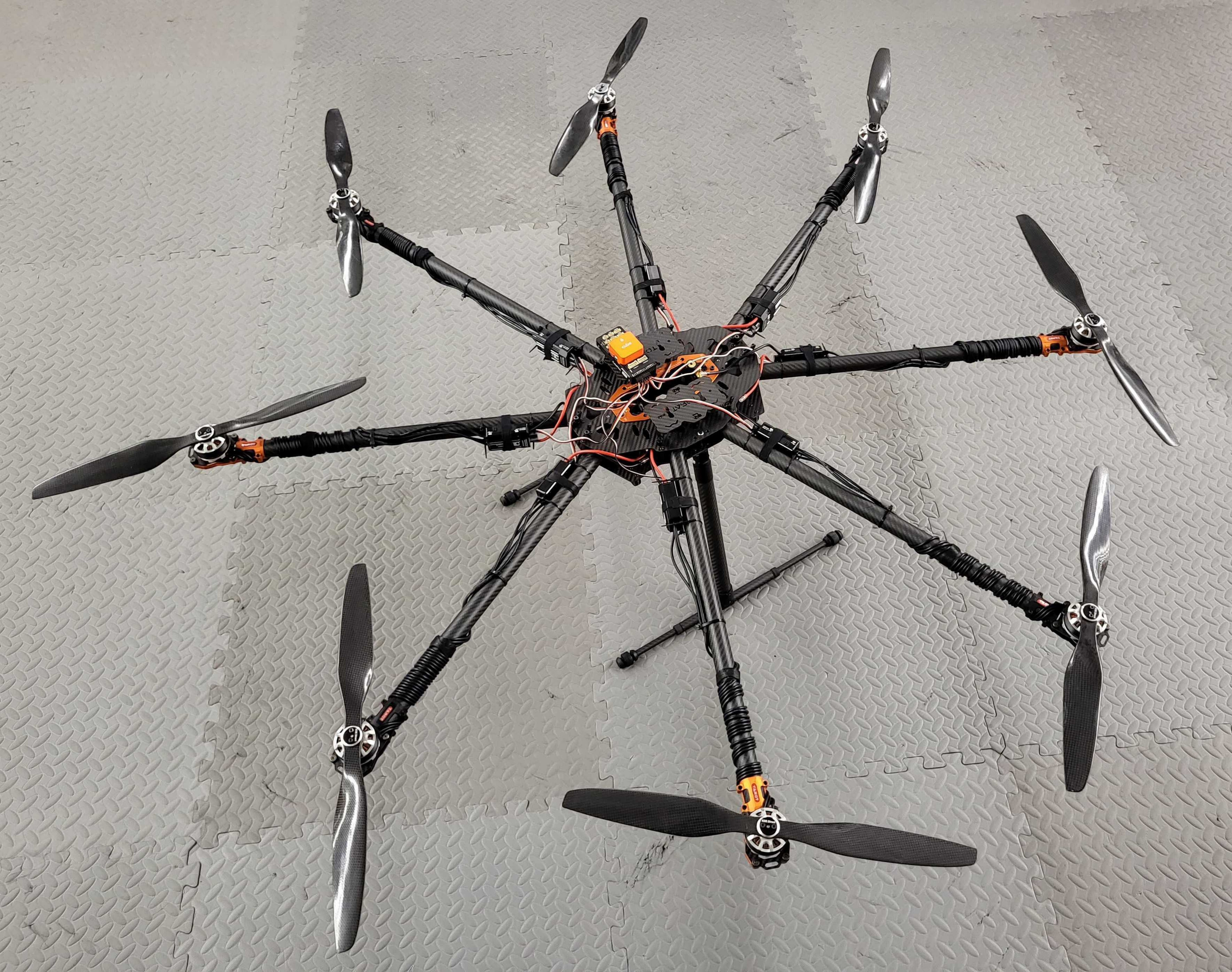}\label{fig:T18} }}
   \caption{(a) Schematic diagram of the electrical components of the octocopter. The Cube Orange (ADS-B Carrier Board) flight controller (\textit{center}) is connected to all components as the main hub. The lipo battery (\textit{center-left}) provides the power to run all components. The electronic speed controller (\textbf{center-right)} receives the control inputs from the Cube to regulate the brushless motor (\textit{top-right}) speed. The GPS module (\textit{top-left}) sends the position information signals to the Cube; then, it transfers the information to GCS. (\textit{bottom-left}). The RC controller (\textit{bottom-right}) will be used in the emergency situation to manually control the octocopter. (b) Tarot T$18$ octocopter, built to validate our proposed controllers.}
   \label{fig:Setup}
\end{figure}

MC-dropout \cite{mcdropout} will be used to compute the model uncertainty or the DRL policy variance. Dropout \cite{dropout} is traditionally used for regularizing networks. It randomly deactivates network units in each forward pass by multiplying the unit weights with a dropout mask. The dropout mask is a set of Bernoulli random variables of value $[0,1]$, each with a probability $p$. Traditionally, dropout is deactivated during a test, and each unit is multiplied with $p$. However, \cite{mcdropout} has shown that activation of dropout during a test named MC-dropout gives model uncertainty estimates by approximating Bayesian inference in deep Gaussian processes. To retrieve the model uncertainty with dropout, in this proposed effort, we will execute multiple forward passes for the DRL network with different dropout masks and acquire a DRL policy distribution. Data augmentation (DA) \cite{data_aug} is a concept that analyzes the model uncertainty in runtime by querying the model multiple times with slightly disturbed inputs. The uncertainty is estimated by measuring how diverse the recommended actions for a given test scenario are. This method is typically used for image processing and helps to improve the image generalization model performance via combining the outputs of several transformed versions of a test image \cite{matsunaga2017image}. Here, we implement DA on the state information, such as position and velocity. First, we will normalize the inputs, add a Gaussian noise, and finally clip them to be in the range of $0$ to $1$. We generate some augmented examples per case to obtain an action distribution and feed them to the DRL model.

\begin{figure}[h!]
\centerline{\includegraphics[width=0.8\textwidth]{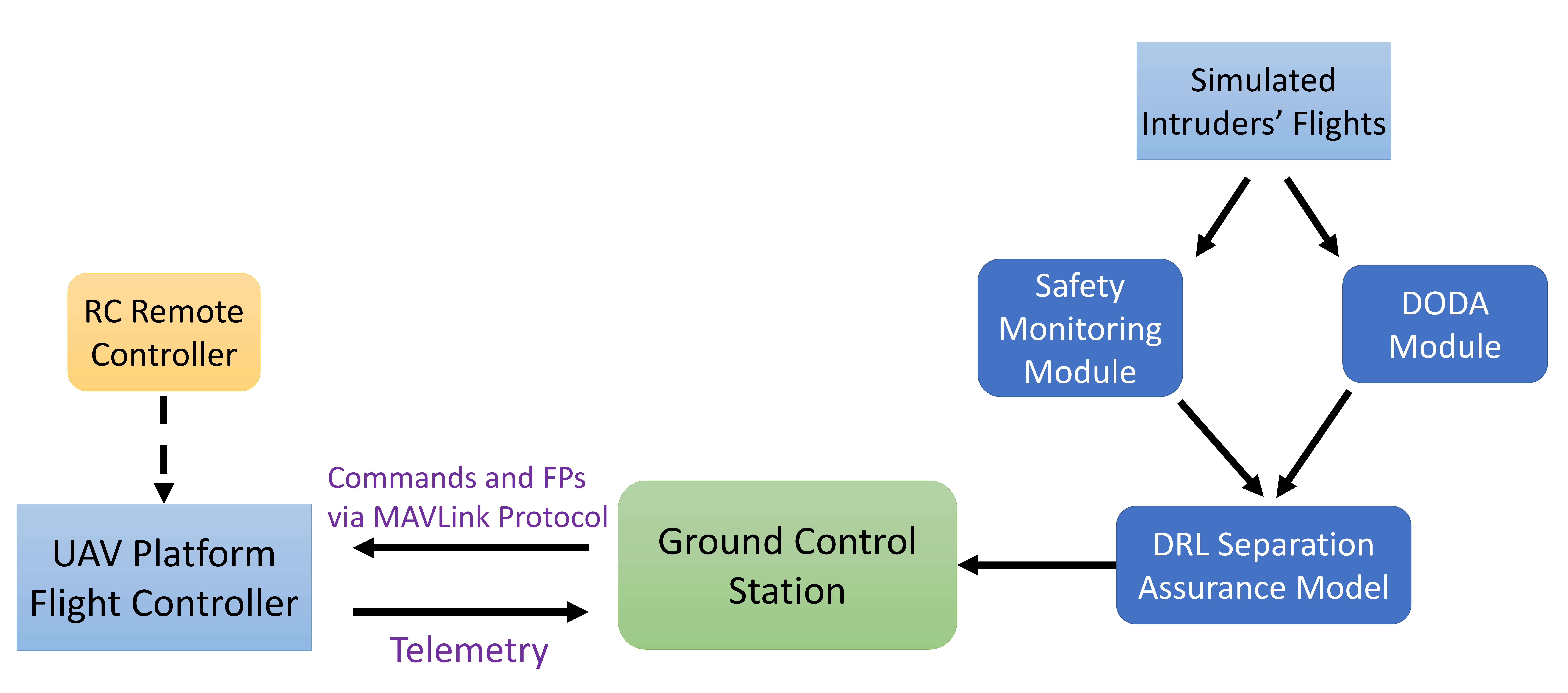}}
\caption{The integration between online safety verification models and hardware testbed. The runtime safety monitoring and bootstrapping modules are implemented in the ground station initially. The flight information of the simulated intruders are stored and played within the real UAS is flown in the airspace. The flight controller communicates with GCS to receive the control commands. The RC remote controller plays a backup role to avoid undesired and unpredictable motion.}
\label{fig:harwareintegration}
\end{figure}

\subsection{Hardware Testbed, Integration, and Demonstration Plan}

The scientific and commercial spheres are paying increasing attention to multirotor vehicles and systems with more than two rotors (e.g., octocopters). Multirotors are very maneuverable and robust due to their greater number of rotors. Thus, we select an octocopter to validate our proposed control strategies and demonstrate a safety verification framework for learning-based aviation on actual experimental hardware.

\paragraph{Hardware setup.}
A UAV platform (see Fig.~\ref{fig:T18}), a foldable and lightweight Tarot T$18$ octocopter, is built and developed to evaluate and demonstrate our proposed methodologies. The specific dimensions of the octocopter's frame are: 
\begin{itemize}
    \item Arms diameter :  $25$mm
    \item Motor to motor distance : $1,200$mm
    \item Height : $380$mm
    \item Body size : $250$mm $\times$ $240$mm
\end{itemize}

The schematic diagram of the primary components of the experimental setup is shown in Fig. \ref{fig:SetupScheme}. The UAS platform is equipped with eight $1125$ W brushless motors (KDE$4213$XF-$360$) connected to a UHV electronic speed controller (KDE-UAS$40$UVC). Eight Tarot $1855$ carbon propellers are attached to the motors. Implementing our proposed methods requires a flight controller, a GPS module, and a remote controller (RC). We use the Cube Orange (ADS-B Carrier Board) flight controller. One can find and view detailed information about the octocopter in the mission planner of this module, including altitude, speed, identification, etc. The flight controller allows us to fly the UAS manually with an RC controller (Jumper T$18$ Pro V$2$, multi-protocol RF Module OpenTX Radio with RDC$90$ Gimbals) or automatically with pre-defined waypoints. Also, we use a GNSS module (Ublox Neo-M$8$N GPS with compass), which is compatible with the Cube. It has a navigation update rate of up to $10$ Hz. The position information is estimated using GPS data. GPS data is inserted into the flight control system through the external interface and communicates with the ground station. A $22,000$ mAh-$22.2$ V lipo battery (Lumenier) is used in the octocopter to provide the necessary power for all electronic components. The flight time will be around $20$ minutes when fully instrumented.

\begin{figure}[t!]
    \centering
   \subfloat[case 1: merging]{{\includegraphics[width=7.5cm]{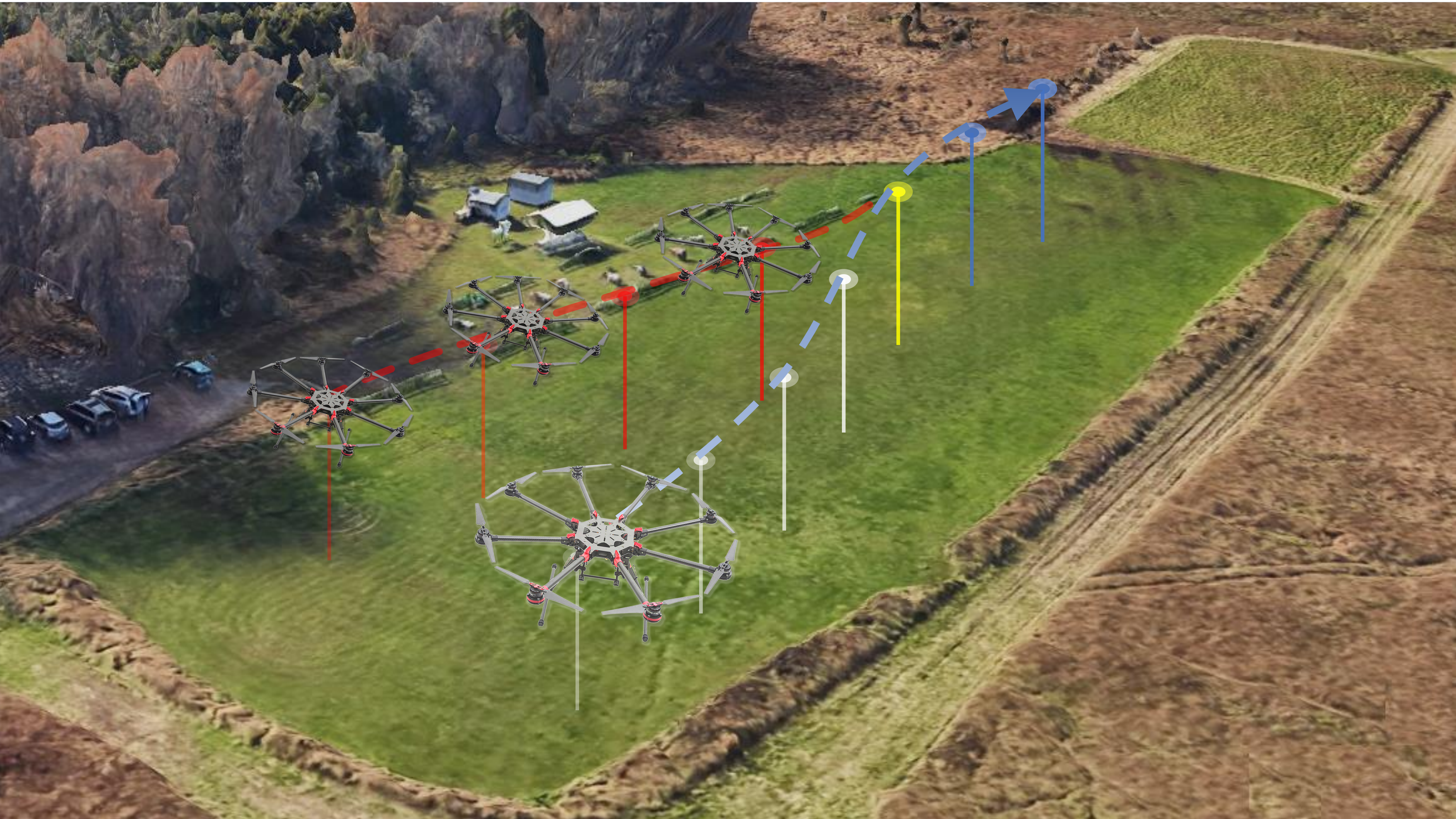}\label{fig:case1} }}
    \qquad
   \subfloat[Case 2: intersections]{{\includegraphics[width=7.5cm]{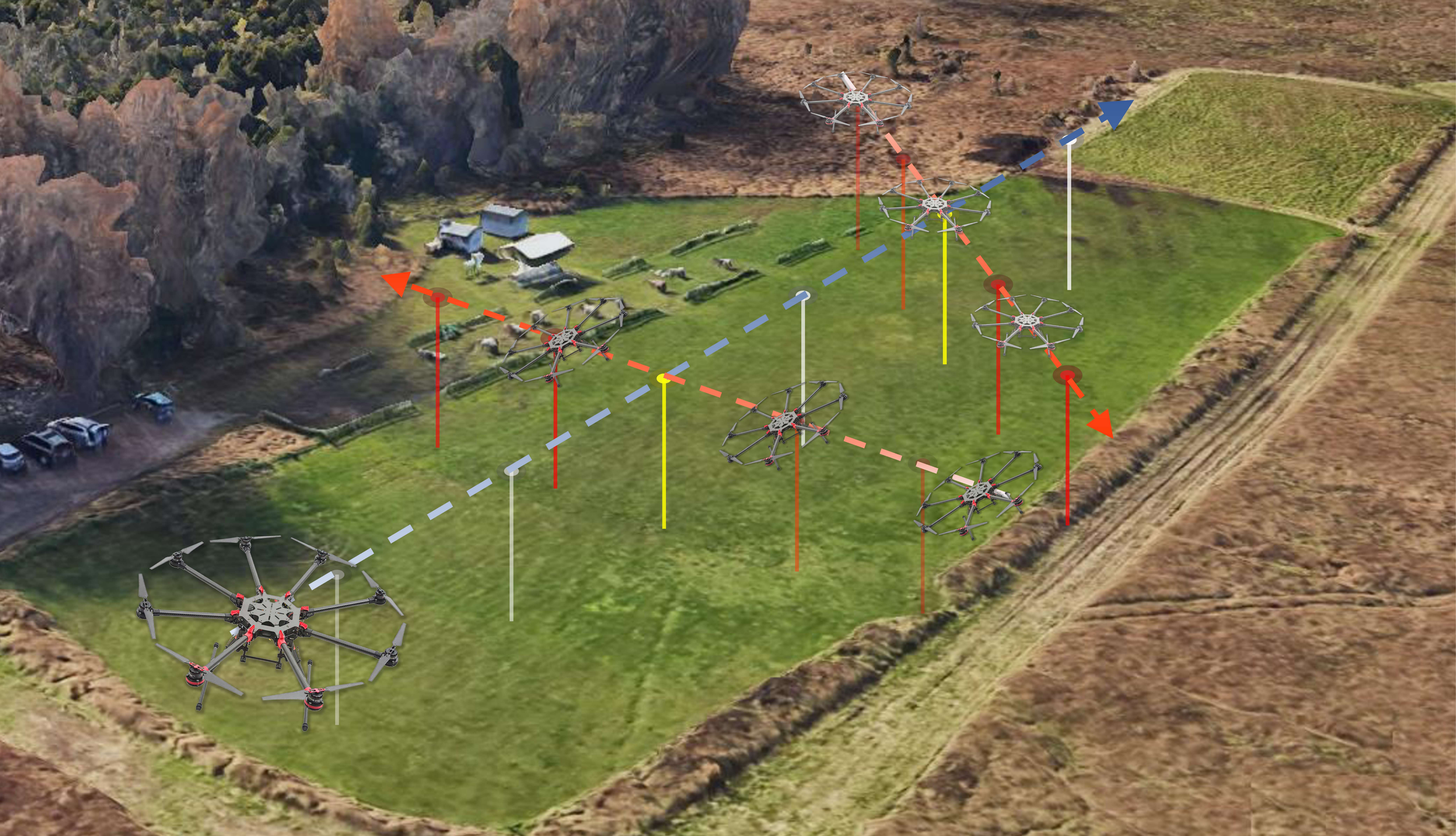}\label{fig:case2} }}
   \caption{Two selected collision avoidance use cases. (a) Merging route; the merging point is indicated using a yellow marker. The ownship aircraft is on the dashed gradient blue line, and the intruders are on the dashed red lines. (b) Intersections of flight routes; the intersections are yellow markers. Dashed gradient red lines show the other two flight routes. The ownship should adjust its speed to avoid a collision at intersections. The waypoints of the ownship and intruders are indicated by white and red markers, respectively.}
   \label{fig:UseCas}
\end{figure}

\paragraph{Hardware integration and demonstration plan.}
Since the collision avoidance flight tests may result in damage to the UAS equipment, the selected UAV and the use case settings will be carefully examined in simulations. We consider the experimental setup as the ownship and other UAVs as intruders, which are virtually simulated in advance. A high-level illustration of the online control process is presented in Fig.~\ref{fig:harwareintegration}. First, the intruders are simulated. Their time-dependent flight information, such as position and velocity, is stored in the ground control station (GCS)—the online reachability and bootstrapping-based models as our proposed safety models are implemented in the GCS. The ownership flight path is planned, and the desired waypoints are sent to the flight controller (Cube see Fig.~\ref{fig:SetupScheme}) via communication through the GCS. As the ownship is flown, the intruder's flight data is played simultaneously, such that the ownship's flight route has intersections with the simulated traffic. Based on the executed information, the DRL safety models avoid a collision between the ownship and the intruders by changing the velocity of the ownship. The DRL control commands will be piped into the flight Cube from GCS.

We will conduct the flight tests, focusing on verifying the DRL model in aircraft collision avoidance use cases for static and dynamic obstacles. Figure~\ref{fig:UseCas} shows two selected flight tests (merging and intersections) and the actual outdoor flight test field (the Northern Virginia Radio Control (NVRC) Poplar Ford Park Field, which is a model airplane field located in Centreville, Virginia). Figure~\ref{fig:case1} shows a merging scenario. The ownship will merge its fight path with intruders at the yellow marker. The speed and the distance between intruders are fixed at their fight path (dashed red line). The ownship on the dashed gradient blue line should adjust the flight speed on its waypoints (white markers) to safely merge into the main flight route. Figure~\ref{fig:case2} also demonstrates an intersection scenario in which we have two fixed routes with many flying intruders at a fixed speed. The ownship should adjust its speed to avoid collisions at the intersections. In the next step, to verify the robustness of our proposed control scheme, we try to randomly change the speeds of the intruders on-the-fly or make them vary at varying speeds instead of fixed ones.

\section{Conclusion}
We proposed a framework for design-time and operational assurance of learning-based components in safety-critical aviation systems. From the design-time assurance standpoint, we proposed a suite of mixed-fidelity methods that coordinate and integrate knowledge from various fidelity testing environments. Efforts include developing rigorous mathematical guarantees to set up simulations across multiple fidelity levels by combining formal logic with multi-fidelity optimization and reinforcement learning techniques. From the run-time assurance point of view, we presented an online safety guard to detect and reduce unsafe actions by combining reachability-based trajectory monitoring and statistics-based model uncertainty estimation. The dynamics-based module provides runtime monitoring for potential unsafe actions, as assigned by a learning-based model; should the recommended action/decision exhibit precursor behaviors that potentially cause the vehicle's trajectory to reach an unsafe state or violate a safety specification, the monitoring module will alert the vehicle control (or switch to a backup controller) to resort to a safe action. We believe our proposed framework offers feasible solutions for the safety verification of learning-based components and could provide the FAA with a viable path to certify learning-based models in avionics systems.

\section{Acknowledgment}
The paper is disseminated under the sponsorship of the U.S. Department of Transportation in the interest of information exchange. The U.S. Government assumes no liability for the contents or use thereof. The U.S. Government does not endorse products or manufacturers. Trade or manufacturers’ names appear herein solely because they are considered essential to the objective of this paper. The findings and conclusions are those of the authors and do not necessarily represent the views of the funding agency. This document does not constitute FAA policy. This research was supported by the FAA under contract No. 692M15-21-T-00022.

\bibliography{peng.bib,hao.bib,ref.bib, stress_test.bib}

\end{document}